\documentclass[article]{JHEP3} % 10pt is ignored!

\JHEPspecialurl{http://jhep.sissa.it/JOURNAL/JHEP3.tar.gz}

\usepackage{epsfig}
\usepackage{graphicx}
\usepackage{subfigure}
\usepackage{amsmath}
\usepackage{multirow}

\newcommand{\bea}{\begin{eqnarray}}
\newcommand{\eea}{\end{eqnarray}}
\newcommand{\nl}{\nonumber\\}

\newcommand\fverbdo{\egroup\medskip\noindent%
                        \fbox{\unhbox\pippobox}\ }
\newcommand\fverbit{\egroup\item[\fbox{\unhbox\pippobox}]}
\newbox\pippobox
\title{Phenomenology of $U(1)_{L_\mu - L_\tau}$ charged dark matter
at PAMELA/FERMI and colliders}

\author{Seungwon~Baek\\
        Institute of Basic Science and
        Department of Physics, Korea University, \\
        Seoul, 136--701, Korea\\
        E-mail: \email{sbaek@korea.ac.kr}}

\author{Pyungwon~Ko\\
        School of Physics, KIAS, \\ Seoul, 130--722, Korea\\
        E-mail: \email{pko@kias.re.kr}}

\received{February xx, 200x}            %%
\revised{}
\accepted{November 27, 2001}            %% These are for published papers.

\abstract{
Recent data on $e^+/e^-$ and $\bar{p}$ cosmic rays suggest that
dark matter annihilate into the standard model (SM) particles
through new leptophilic interaction.
In this paper, we consider a standard model extension with the gauged
$U(1)_{L_\mu - L_\tau}$ group, with a new Dirac fermion  charged
under this $U(1)$ as a dark matter.
We study the muon $(g-2)_\mu$, thermal relic density of the cold
dark matter, and the collider signatures of this model.
$Z^{'}$ productions at the Tevatron or the LHC could be easily order of
$O(1)-O(10^3)$ fb.
 }


\dedicated{}

\begin{document}

%\maketitle  IS IGNORED %%%%%%%%%%%

\section{Introduction}

The Standard Model (SM) should be extended in order to accommodate the nonbaryonic cold dark matter (CDM) of the universe.
There are many options for cold dark matter candidates in particle
physics models: axion, neutralino or gravitino lightest supersymmetric 
particle (LSP) \cite{lsp,bertone}, lightest Kaluza-Klein particle (LKP) \cite{bertone}, 
and lightest particle in a hidden sector \cite{hidden}, to name a few.
However, we do not have enough information on the detailed nature of CDM,
except that $\Omega_{CDM} h^2 =  0.106 \pm 0.008$~\cite{PDG}.
More information from direct and indirect dark matter searches and 
colliders are indispensable for us to diagnose the particle identity of 
the CDM, 
namely its mass and spin and other internal quantum number(s).

Recently, PAMELA reported a sharp increase of positron fraction 
$e^+/(e^+ + e^-)$
in the cosmic radiation for the energy range 10 GeV to 100 GeV~\cite{PAMELA_e},
and no excess in $\bar{p}/p$~\cite{PAMELA_p} from the theoretical
calculations. And also very recently, Fermi-LAT~\cite{Fermi-LAT}
and HESS~\cite{HESS} data showed clear excess
of the $e^+ + e^-$ spectra in the multi-hundred GeV
range above the conventional model~\cite{background},
although they do not confirm the previous ATIC~\cite{ATIC} peak.

These data (at least a part of them) could be due to some astrophysical 
origins such as pulsars~\cite{Hooper:2008kg,Profumo:2008ms}. 
A more exciting possibility from particle physics point of view 
would be interpreting them as indirect signatures of cold dark matters 
through their pair annihilations or decays. 
In this paper we take the second avenue, namely particle physics explanations of positron excess.
Model independent study\cite{model-indep-study} show that 
multi-TeV scale DMs dominantly annihilating into SM leptons, 
especially into $\tau^+ \tau^-$ or 4 $\mu$'s, are most favored.
If one assumes the standard cosmology, large boost factor (BF) of 
${\cal O}(10^3)$ is also required to enhance the event rates.

From the view point of model building and grand unification, it is non-trivial
to construct this kind of leptophilic model. 
However it is still phenomenologically
viable, and we have to verify or falsify this class of models
by comparing the predictions with the data.
In this paper we explicitly construct a leptophilic model
 and work out the physical consequences in detail.
There are already many papers available studying the implications of 
the PAMELA/FERMI
data in different models and/or context~\cite{DM-theory}.

The simplest model for the leptophilic (or hadrophobic) gauge interaction is
to gauge the global $U (1)_{L_{\mu} - L_{\tau}}$ symmetry of the standard model
(SM), which is anomaly free \cite{He:1990pn,He:1991qd,Foot:1994vd,baek_he_ko}.
Within the SM, there are four global $U(1)$ symmetries which are anomaly free:
\[
L_e - L_\mu , \ \  L_\mu - L_\tau , \ \ L_\tau - L_e , \ \  B - L
\]
One of these can be implemented to a local symmetry without anomaly.
The most popular is the $U(1)_{B-L}$, which can be easily implemented
to grand unified theory. Two other symmetry involving $L_e$ are tightly
constrained by low energy and collider data.
On the other hand, the $L_\mu - L_\tau$ symmetry is not so tightly
constrained, and detailed phenomenological study has not been 
available yet.  Only the muon $(g-2)_\mu$ and the phenomenology
at muon colliders have been discussed~
\cite{baek_he_ko,Gninenko:2001hx}. This model can be extended 
by introducing three right-handed neutrinos and generate the neutrino 
masses and mixings via seesaw mechanism \cite{He:1991qd}.  Also 
$U(1)_{L_\mu - L_\tau}$  can be embedded into a horizontal $SU(2)_H$
\cite{He:1991qd} acting on three lepton generations, which may 
be related with some grand unification.

In this paper, we extend the existing $U(1)_{L_\mu - L_\tau}$ model 
by including a complex scalar $\phi$ and a spin-1/2 Dirac fermion 
$\psi_D$,  with $U(1)_{L_{\mu} - L_{\tau}}$ charge 1.
There is no anomaly regenerated in this case, since we introduced 
a vectorlike fermion. The complex scalar $\phi$  gives a mass 
to the extra $Z^{'}$ by ordinary Higgs mechanism. And the Dirac fermion
$\psi_D$ plays a role of the cold dark matter, whose pair annihilation 
into $\mu$ or $\tau$ explains the excess of $e^+$ and no $\bar{p}$ 
excess as reported by PAMELA~\cite{PAMELA_e,PAMELA_p}. 
Then we study CDM cosmology (thermal relic density and (in)direct  signatures) 
and collider phenomenology of the 
$U(1)_{L_\mu - L_\tau}$ model with Dirac fermion dark matter in detail. 

In Sec.~2, we define the model and discuss the muon $(g-2)_\mu$ 
in our model.  
In Sec.~3, we calculate the thermal relic density of the CDM $\psi_D$, 
and identify the parameter region that is consistent with the data 
from cosmological observations.  We also present the signatures for 
indirect search experiments: $e^+ e^-$, neutrinos,
and gamma rays from the DM annihilations, including the Sommerfeld enhancement.
In Sec.~4, we study the collider signatures of the model at various 
colliders encompassing Tevatron, B factories, LEP(2), the $Z^0$ pole 
and LHC, including productions and decays of $Z^{'}$, the SM Higgs 
boson and the newly introduced $U(1)_{L_\mu - L_\tau}$ charged 
scalar boson.   Our results are summarized in Sec.~5. 
We note that this model was discussed briefly by Cirelli, {\it et.al} in 
Ref.~\cite{model-indep-study} in the context of the muon $(g-2)_\mu$ 
and the relic density. In this paper, we present the quantitative analysis 
on these subjects in detail, as well as study various signatures at 
various colliders including the Tevatron and the LHC.

\section{Model and the muon $(g-2)_\mu$}

The new gauge symmetry $U(1)_{L_\mu - L_\tau}$ affects only the 2nd and
the 3rd generations of leptons. We assume $l_L^{i=2(3)}$, $l_R^{i=2(3)}$
($i$: the generation index) carry $Y^{'} = 1(-1)$.
We further introduce a complex scalar $\phi$ with $(1,1,0)(1)$ and
a Dirac fermion $\psi_D$ with $(1,1,0)(1)$, where the first and the second
parentheses represent the SM and the $U(1)_{L_\mu - L_\tau}$ quantum numbers of $\phi$ and $\psi_D$, respectively. 
The covariant derivative is defined as
\begin{equation}
  D_\mu = \partial_\mu + i e Q A_\mu + i {e \over s_W c_W}
( I_3 - s_W^2 Q ) Z_\mu + i g^{'} Y^{'} Z^{'}_\mu
\end{equation}
The model lagrangian is given by \footnote{Similar idea for the DM 
was considered in~\cite{Feldman:2007wj,Cheung:2007ut}
in the context of Stueckelberg $U(1)_X$ extension of the SM model.}
\begin{eqnarray}
  {\cal L}_{\rm Model} & = & {\cal L}_{\rm SM} + {\cal L}_{\rm New}
\\
\label{eq:L_new}
{\cal L}_{\rm New} & = & - {1\over 4} Z^{'}_{\mu\nu} Z^{'\mu\nu}
+ \overline{\psi_D} \left(i D \cdot \gamma -M_{\psi_D} \right) \psi_D \nl
 & & + D_\mu \phi^* D^\mu \phi - \mu_\phi^2 \phi^* \phi
- \lambda_\phi ( \phi^* \phi)^2  
- \lambda_{H\phi} \phi^* \phi H^\dagger H.
\end{eqnarray}
In general, we have to include renormalizable kinetic mixing term for 
$U(1)_Y$ and $U(1)_{L_\mu - L_\tau}$ gauge fields, $B_{\mu\nu} Z^{' \mu\nu}$, which will lead 
to the mixing between $Z$ and $Z^{'}$. Then the dark matter pair 
can annihilate into quarks through $Z-Z^{'}$ mixing in our case, and  
the $\bar{p}$ flux will be somewhat enhanced, depending on  the size 
the $Z-Z^{'}$ mixing.  However, electroweak precision data and collider 
experiments give a strong constraint on the possible mixing parameter, 
since the mixing induces the $Z^{'}$ coupling to the quark sector. 
Furthermore, if one assumes that the new $U(1)_{L_\mu - L_\tau}$ is 
embedded into a nonabelian gauge group such as $SU(2)_H$ or $SU(3)_H$, 
then the kinetic mixing term is forbidden by this nonabelian gauge 
symmetry \cite{He:1991qd}.  In this paper, 
we will assume that the kinetic mixing is zero to simplify 
the discussion and to maximize the contrast between the positron
and the antiproton fluxes from the dark matter annihilations. 
\footnote{Because of this simplification, the direct detection rate from 
the CDM (in)elastic scattering off nuclei vanishes identically. However 
this is no longer true if the kinetic mixing is included. 
This case is discussed in brief in Sec.~3.2.}
   
In this model, there are two phases for the extra $U(1)_{L_\mu - L_\tau}$
gauge symmetry depending on the sign of $\mu_\phi^2$ :
\begin{itemize}
\item Unbroken phase: exact with $\langle \phi \rangle = 0$,
$\mu_\phi^2 > 0$ and $M_{Z^{'}} = 0$,
\item Spontaneously broken phase: by $\mu_\phi^2 < 0$, nonzero
$\langle \phi \rangle \equiv v_\phi \neq 0$, and $M_{Z^{'}} \neq 0$
\end{itemize}
In the unbroken phase, the massless $Z^{'}$ contribute to the muon $(g-2)_\mu$
as in QED up to the overall coupling:
\begin{equation}
\Delta a_\mu = {\alpha^{'} \over 2 \pi} .
\label{eq:amu_massless}
\end{equation}
Currently there is about 3.4$\sigma$ difference between
the BNL data~\cite{BNL} and the SM predictions~\cite{g-2_theory} in $(g-2)_\mu$:
\begin{equation}
 \Delta a_\mu = a_\mu^{\rm exp} - a_\mu^{\rm SM} = (302 \pm 88) \times 10^{-11}.
\end{equation}
The $\Delta a_\mu$ in (\ref{eq:amu_massless}) can explain this discrepancy,
if $  \alpha^{'} \sim 2 \times 10^{-8}$. However, this coupling is too small 
for the thermal relic density to satisfy the WMAP data. The resulting
relic density is too high by a several orders of magnitude.
Also the collider signatures will be highly suppressed. 
Therefore we do not consider this possibility any further, and consider
the massive $Z^{'}$ case (broken phase) in the following.

In the broken phase, it is straightforward to calculate the $Z^{'}$
contribution to $\Delta a_\mu$. We use the result obtained in 
Ref.~\cite{baek_he_ko}:
\begin{equation}
\Delta a_\mu = {\alpha^{'} \over 2 \pi}~\int_0^1 dx { 2 m_\mu^2 x^2 (1-x)
\over x^2 m_\mu^2 + ( 1-x) M_{Z^{'}}^2 } \approx
{\alpha^{'} \over 2 \pi} {2 m_\mu^2 \over 3 M_{Z^{'}}^2 }
\end{equation}
The second approximate formula holds for $m_\mu \ll  M_{Z^{'}}$.
In Fig.~1, shown in the blue band is the allowed region of
$M_{Z^{'}}$ and $\alpha^{'}$ which is consistent with the BNL data
on the muon $(g-2)_\mu$ within 3 $\sigma$ range.
There is an ample parameter space where the discrepancy between 
the BNL data and the SM prediction can be explained within the model.

\section{Dark matter : Relic density and (In)direct signatures}

\subsection{Thermal relic density}

In our model, the Dirac fermion $\psi_D$ and its antiparticle 
$\overline{\psi}_D$ are  CDM candidates.
The thermal relic density of $\psi_D$ and $\overline{\psi}_D$ 
is achieved through the DM annihilations into muon, tau leptons 
or their neutrinos through s-channel $Z'$-exchange. They can also
annihilate into the real $Z'$ pairs when kinematically allowed.
\begin{eqnarray}
\psi_D \bar{\psi}_D &\rightarrow& Z^{'*} \rightarrow l^+ l^- ,
\nu_l \bar{\nu}_l  \quad (l = \mu, \tau), \nonumber\\
\psi_D \bar{\psi}_D &\rightarrow& Z^{'} Z^{'}.
\end{eqnarray}
We modified the micrOMEGAs~\cite{micromegas} in order to calculate
the relic density of the $U(1)_{L_\mu - L_\tau}$ charged $\psi_D$ CDM.
It is easy to fulfil the WMAP data on $\Omega_{\rm CDM}$ for a wide
range of the DM mass, as shown in Fig.~\ref{fig:relic}. The black curves
represent constant contours of $\Omega h^2 = 0.106$ in the 
$(M_{Z'},\alpha)$-plane for $M_{\psi_D} = 10, 100, 1000$ GeV (from below).  
We can clearly see  the $s-$channel resonance effect of 
$Z^{'} \rightarrow \psi_D \bar{\psi}_D$ near $M_{Z'} \approx 2 M_{\psi_D}$.
The blue band is the allowed region by the $(g-2)_\mu$ at the 3 
$\sigma$ level.
We also show the contours for the $Z'$ production cross sections  
at various colliders:  B factories ($1 {\rm fb}$, red dotted), Tevatron 
($10 {\rm fb}$, green dot-dashed), LEP($10 {\rm fb}$, pink dotted),  LEP2($10 {\rm fb}$, orange dotted) and LHC ($1$ fb, $10$ fb and $100$ fb in  blue dashed curves).
The cross sections in the parentheses except the LHC case roughly 
correspond to the upper bounds that each machine
gives. Therefore the left-hand sides of each curve is ruled out by the
current  collider data.
Note that a larger parameter space can be accessed by the LHC.
These issues and other collider siugnatures are covered in the next 
section.

The current experimental mass bound of SM-like $Z'$ is 923 GeV from the search
for a narrow resonance in electron-positron events~\cite{CDF-Zprime}.
We emphasize, however, that in our model the $Z'$ boson 
as light as $\sim 10$ GeV is still
allowed by present data from various colliders. 
It is mainly because the production cross section at the Tevatron is
suppressed since $Z'$ should be produced from the couplings to the 
2nd and 3rd family leptons.

In the range $ 100\; {\rm GeV} \lesssim M_{\psi_D} \lesssim 10\; {\rm TeV}$,
$\alpha \gtrsim 10^{-3}$ and $100 \; {\rm GeV} \lesssim M_{Z'} \lesssim 1 $ TeV,
the relic density and $\Delta a_\mu$ constraints can be easily satisfied 
simultaneously while
escaping the current collider searches. 
We note that if the $(g-2)_\mu$ constraint is not considered seriously
or if we assume there are other sector which saturate the $(g-2)_\mu$ 
upper bound,
then all the region in the right-hand side of the
blue band is also allowed.

\FIGURE{
\epsfig{file=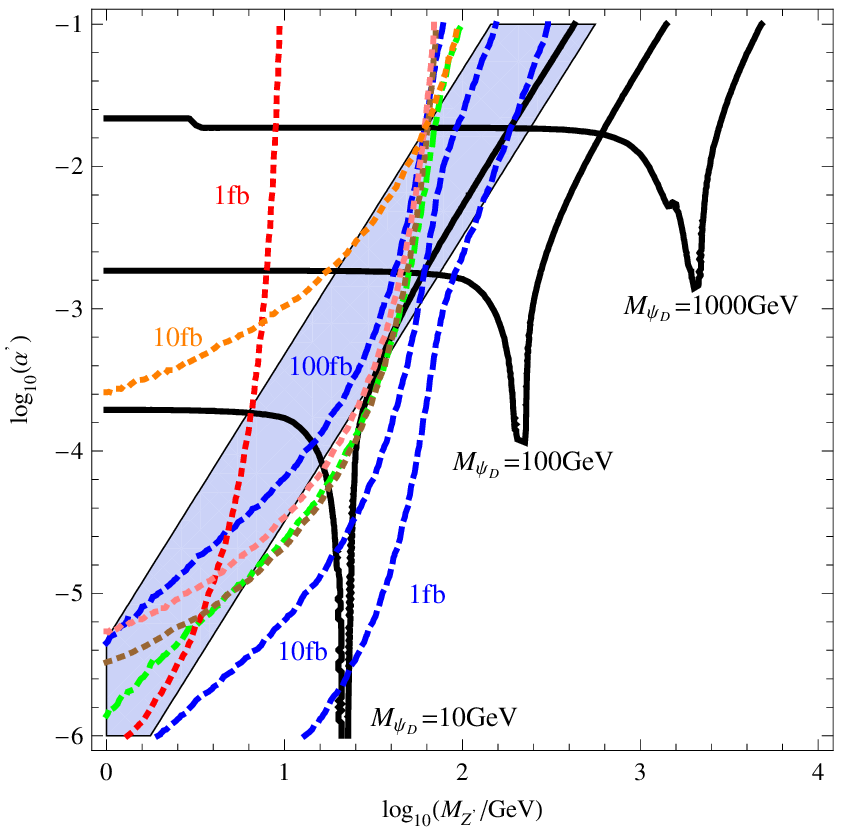,width=12cm,height=9cm}
\caption{
The relic density of CDM (black), the muon $(g-2)_\mu$ (blue band),
the production cross section at $B$ factories (1 fb, red dotted),
Tevatron (10 fb, green dotdashed), LEP (10 fb, pink dotted), LEP2 (10 fb, orange dotted),
LHC (1 fb, 10 fb, 100 fb, blue dashed)
and the $Z^0$ decay width (2.5 $\times 10^{-6}$ GeV, brown dotted) in the
$( \log_{10} \alpha^{'}, \log_{10} M_{Z^{'}} )$ plane. For the relic
density,  we show three contours with $\Omega h^2 = 0.106$ for $M_{\psi_D} $ = 10 GeV, 100 GeV and
1000 GeV. The blue band is allowed by $\Delta a_\mu = (302 \pm 88) \times 10^{-11}$ within
3 $\sigma$.}
\label{fig:relic}
}

\subsection{Direct detection rates}

Since we ignored the kinetic mixing between the new $U(1)$ gauge 
boson and the SM $U(1)_Y$ gauge boson $B_\mu$,  there would be 
no signal in direct DM detection experiments in this model.
The messenger $Z^{'}$ does not interact with electron, quarks or gluons 
inside nucleus.   Also there would be no excess in the antiproton flux in cosmic rays in this case, while one could have an excess in the positron signal in a manner consistent with the PAMELA/Fermi data.
However there would be a small kinetic mixing between two $U(1)$ 
gauge field strength tensor. 
If we assume a small kinetic mixing $\theta (\sim 10^{-3} = 10^{-2})$ between the $Z^{'}_\mu$ and photon,  
then spin-independent cross section for direct detection rate will be 
given by 
\begin{equation}
\sigma^{\rm SI} = \frac{4}{\pi}~\left(  \frac{M_{\psi_D} M_A}{M_{\psi_D} 
+ M_A}  \right)^2 ~\left[ \lambda_p Z + \lambda_n (A-Z) \right]^2 ,
\end{equation}
where $Z$ and $A$ are the atomic number and the mass number of 
a nucleus.  The couplings $\lambda_{p(n)}$'s of the CDM to proton 
and neutron are given by 
\begin{eqnarray}
\lambda_p & = & \pm 
\frac{eg^{'} }{2 M_{Z^{'}}^2}~\left[ \theta_{Z^{'}\gamma}
+ \frac{\theta_{Z^{'}Z}}{4 \sin\theta_w \cos\theta_w}~
(1 - 4 \sin^2\theta_w )
\right]
\\
\lambda_n & = & \mp 
\frac{eg^{'} \theta_{Z^{'}Z}}{8M_{Z^{'}}^2 \sin\theta_w \cos\theta_w}
\end{eqnarray}
where the upper (lower) sign corresponds to $\psi_D (\overline{\psi}_D)$ 
DM scattering. 
Note that $\lambda_p$ and $\lambda_n$ are dominated by the 
$Z^{'}-\gamma$ and $Z^{'}-Z$ mixing, respectively. This is because 
photon couples to the nucleon charge, whereas the $Z^0$ couples 
to the neutral current weak charge of a nucleon. 
The $Z^0$ coupling to a proton is proportional to 
$(1 - 4 \sin^2 \theta_w)$, and thus highly suppressed compared to 
the $Z^0$ coupling to a neutron.  Also let us note that the cross 
section on a nucleus could be small,   if there is a cancellation between 
$\lambda_p$ and $\lambda_n$ terms depending on the sign of 
$\theta_{Z^{'}\gamma}$ and $\theta_{Z^{'}Z}$  in Eq.s~(3.3)-(3.5).

If we consider a dark matter scattering on single proton target, 
one has 
\begin{equation}
\sigma_{\psi_D p}^{\rm SI} %(Z^{'}\gamma \ {\rm mixing})
 \simeq  16 \pi \alpha \alpha^{'} \theta_{Z^{'}\gamma}^2
%\left(  \frac{M_{\psi_D} M_p}{M_{\psi_D} + M_p } \right)^2 
\frac{M_p^2}{M_{Z^{'}}^4}   
%\nonumber 
%\\
 \approx  1.3 \times 10^{-42} cm^2~
\left( \frac{100 \ ({\rm GeV})}{M_Z^{'}}\right)^4~
\left( \frac{\alpha^{'}}{10^{-2}} \right)
\left( \frac{\theta_{Z^{'}\gamma}}{10^{-2}} \right)^2
\end{equation}
which is dominated by $Z^{'}-\gamma$ mixing.  
The resulting cross section is close to the current upper bounds 
from XENON10 \cite{xenon} and CDMS  \cite{cdms} experiments.  
We have assumed that the DM $\psi_D$ is much heavier than proton. 
Similarly, the SI cross section on a neutron target is given by 
\begin{equation}
\sigma_{\psi_D n}^{\rm SI} %(Z^{'}Z \ {\rm mixing}) 
 \simeq \frac{\pi \alpha \alpha^{'}\theta_{Z^{'}Z}^2}{
\sin^2\theta_w \cos^2 \theta_w} 
%\left(  \frac{M_{\psi_D} M_p}{M_{\psi_D} + M_p } \right)^2 
\frac{M_n^2}{M_{Z^{'}}^4}   
%\nonumber  \\
 \approx  4.6 \times 10^{-43} cm^2~
\left( \frac{100 \ ({\rm GeV})}{M_Z^{'}}\right)^4~
\left( \frac{\alpha^{'}}{10^{-2}} \right)
\left( \frac{\theta_{Z^{'}Z}}{10^{-2}} \right)^2
\end{equation}
The scattering cross section on the proton by $Z-Z^{'}$ mixing is suppressed by $(1-4 \sin^2 \theta_w )^2$ relative 
to the scattering cross section on the neutron target, and thus negligible.
In either case, one can evade the bounds from XENON10 and CDMSII
by taking a heavier $Z^{'}$ mass, smaller coupling $\alpha^{'}$ or 
smaller mixing angle $\theta$. If $Z^{'}$ is light, one may have 
too large SI cross section,  in conflict with XENON10 and CDMSII.

\subsection{Sommerfeld enhancement and boost factor (BF)}

The DM annihilation cross section at the freeze-out temperature is typically
$\langle \sigma v \rangle \sim 3 \times 10^{-26}\; {\rm cm^3/sec}$.
To explain the PAMELA or Fermi data, 
$\langle \sigma v \rangle \sim 3 \times 10^{-23}\; {\rm cm^3/sec}$
is required.  Therefore we need a boost factor (BF) of order of $10^3$.
A large BF can come from the so-called Sommerfeld enhancement 
in the $S-$wave DM annihilation. 
The Sommerfeld factor, given by the ratio of  the radial wavefunction 
at infinity to that  at the origin, 
\bea
  S_k = \left | \chi_k(\infty) \over \chi_k(0) \right|^2
\eea
can be calculated by
solving the radial $S-$wave Schr\"odinger equation~\cite{sommerfeld}
with the attractive Yukawa potential in our model:
\bea
-{1 \over 2 M_{\psi_D}} \frac{d^2}{dr^2} \chi_k(r) -{\alpha' \over 2 r} 
e^{-M_{Z'} r}  \chi_k(r)
= {k^2 \over 2 M_{\psi_D}} \chi_k(r).
\eea
Here $k = M_{\psi_D} v$ and $v$ is the relative velocity of two annihilating
DM particles.
The boundary condition for the above Schr\"odinger equation
is
\bea
 \chi'_k(r) \to i k \chi_k(r) \text{ as } r\to \infty, \quad \chi_k(0)=1.
\eea
Fig.~\ref{fig:sommerfeld} shows the prediction for the Sommerfeld enhancement factor in our model for various values of DM:
$M_{\psi_D} = 10, 100, 1000, 2000$ GeV.
For a given values of the DM mass and the $M_{Z'}$,  the $\alpha'$ are chosen in such a way that they satisfy the relic density, {\it i.e.}
each lines are predictions of the Sommerfeld enhancement factor along the constant  relic density contours in Fig.~2.
We can see that it is easy to get the enhancement factor $\sim 10^3$. 

\FIGURE{
\epsfig{file=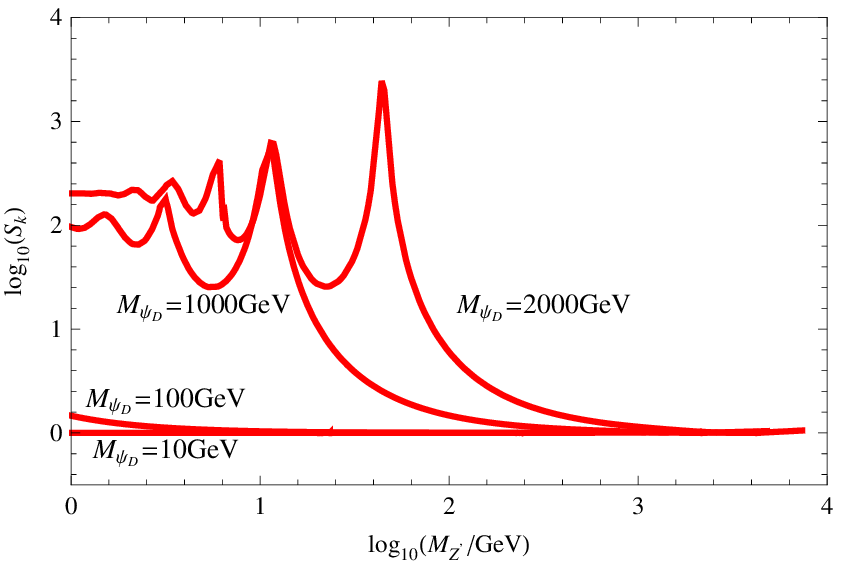,width=12cm,height=9cm}
\caption{
The predictions of Sommerfeld enhancement factor in our model along the constant
contours in Fig.~2.
}
\label{fig:sommerfeld}
}

\subsection{Indirect signatures: positron, neutrino and photon fluxes}

Now we show the prediction of $e^+/(e^+ + e^-)$ spectra in our model 
in Fig.~\ref{fig:PAMELA}. The positron flux at the Earth can be calculated from
the solution of the diffusion equation as~\cite{delahaye}
\bea
  \Phi_{e^+}(\odot,\epsilon) &=& {\beta_{e^+} \over 4 \pi} 
  {\kappa \tau_E \over \epsilon^2}
  \int_\epsilon^\infty d\epsilon_S f(\epsilon_S) \tilde{I}(\lambda_D),
\eea
where $\odot$ represent the position of the sun, $\epsilon = E_{e^+}/E_0$ ($E_0=$ 1 GeV).
The function $f(\epsilon_S)$ is the positron energy spectrum from the dark matter
annihilation at the source.
The ``diffusive halo function"  $\tilde{I}(\lambda_D)$ 
with $\lambda_D^2= 4 K_E \tau_E (\epsilon^{\delta-1}-\epsilon_S^{\delta-1})/(1-\delta)$  
encodes the information on the propagation of positron from the source 
to the Earth.
The parameter 
$\kappa = \eta \langle \sigma v \rangle \left(\rho_\odot/M_{\psi_D}\right)^2$ 
is a factor relevant to the particle physics [$\eta=1/2 ~(1/4)$ for 
Majorana (Dirac) DM].  
The other parameters are: the speed of the positron $\beta_{e^+}$ and  
and $\tau_E = 10^{16}$ sec.
We used the NFW DM density profile~\cite{NFW}:
\bea
 \rho(r) = \rho_\odot \; \left(r_\odot \over r \right)^\gamma \;
   \left(1+(r_\odot/r_s)^\alpha \over 1+(r/r_s)^\alpha \right)^{(\beta-\gamma)/\alpha},
\eea
where $(\alpha, \beta, \gamma, r_s)=(1, 2, 1, 20 \ {\rm kpc})$ and 
$\rho_\odot = 0.3$ GeV/cm$^3$ is the DM density near the Sun.
To obtained the halo funtion $\tilde{I} (\lambda_D)$, we used 
the method suggested in~\cite{baltz}.
We also used the cosmic ray propagation parameters which correspond to the medium
primary antiproton fluxes~\cite{Donato:2003}: 
$\delta=0.70$, $K_0 = 0.0112 {\rm kpc^2/Myr}$,  and $L=4 {\rm kpc}$.  
For the plot, Fig.~\ref{fig:PAMELA}, we fixed the DM mass 
$M_{\psi_D}=2$ TeV.
The required BF is about 5200, which is a little bit larger than the maximal 
Sommerfeld enhancement in Fig.~\ref{fig:sommerfeld} can give.
However, an additional enhancement factor of about 2--3 can be easily obtained from the  clumpy structure of dark matter density.
We also have checked with other DM masses ranging from 1 TeV to 3 TeV, which can also fit the PAMELA data very well.

Although the PAMELA data alone can be fitted with a wide range of the DM mass~\cite{model-indep-study},
the simultaneous fit including the Fermi and HESS data is non-trivial and gives more strong constraint on the DM mass. 
Fig.~\ref{fig:Fermi} shows a fit to the PAMELA, Fermi and HESS data in our model.
First, we obtained the absolute positron flux from the PAMELA data and 
the known background positron and electron spectrum~\cite{background}. The resulting spectra agree with the Fermi data when we rescale the background by a factor $r \sim 0.7$.
This rescaling also makes all the Fermi data lie above the background.
The HESS data has also been allowed a rescale factor $r_H$.
Now we fitted the three parameters $r, r_H$ and the BF to the data, assuming the data are independent
with each other. For the DM mass $M_{\psi_D} = 2 $ TeV, we obtained an excellent fit
$\chi^2_{\rm min}/d.o.f = 53/50$, $r=0.7$, $r_H=0.84$ and ${\rm BF} =5200$.
For the lighter and heavier DM masses, the fit quality becomes worse.
We have also checked that the isothermal DM profile also gives similar results.
The reason can be traced back to the fact that the positrons we observe comes mainly from the sources not that far from the Sun where the DM density does not differ much for different DM profiles, although the 
NFW profile is far more cuspy than the isothermal profile when 
approaching the Galactic center. 

\FIGURE{
\epsfig{file=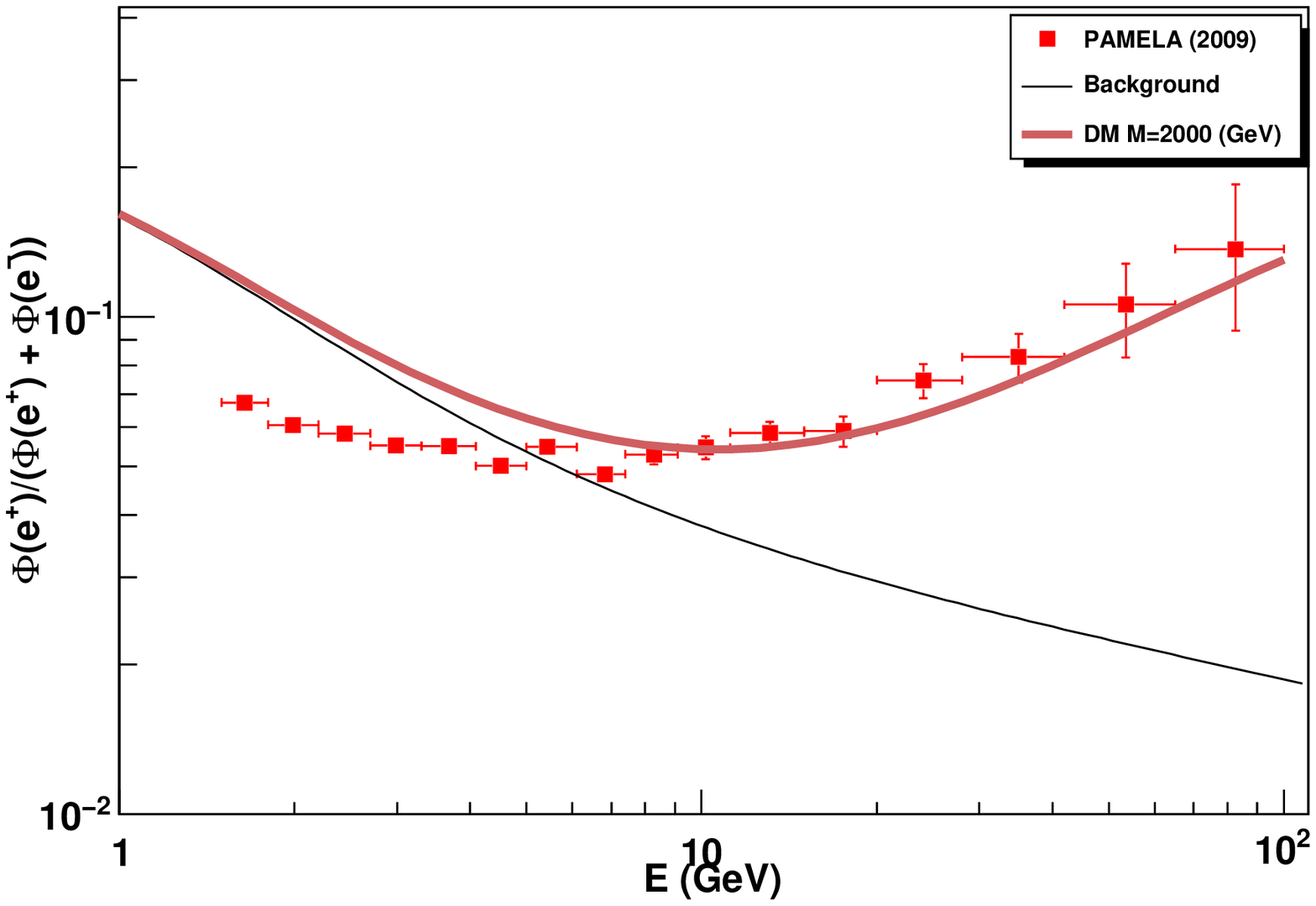,width=12cm,height=9cm}
\caption{
The fit to the PAMELA data in our model.
}
\label{fig:PAMELA}
}

\FIGURE{
\epsfig{file=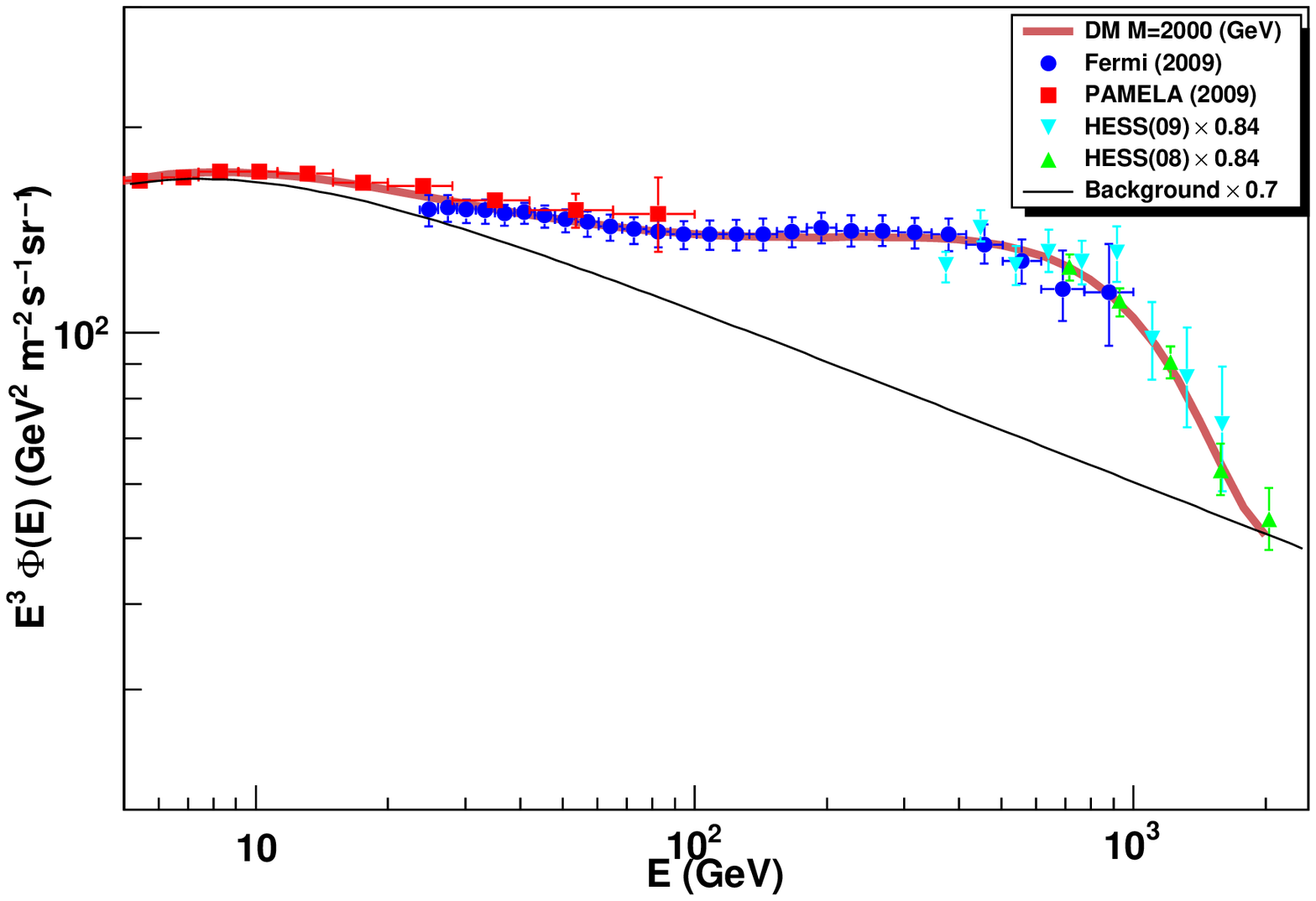,width=12cm,height=9cm}
\caption{
The fit to the PAMELA, Fermi and HESS data in our model.
}
\label{fig:Fermi}
}

Since the DM pair annihilates into the 2nd and 3rd generation leptons including neutrinos in our model, we expect large neutrino flux. 
The neutrino flux can be detected through the upward-going muons 
in the Super-Kamiokande (SK).
Also the neutral pion from the tau decay can produce sizable photon 
flux, which can also be compared with the existing gamma-ray searches 
such as HESS\footnote{See also \cite{Bergstrom:2008ag} for the gamma-ray constraint.}.
The main contribution to the neutrino and the gamma-ray flux comes 
from the Galactic  center where the DM density is the highest. 
For this reason we consider only neutrinos and gamma-rays from the Galactic center.  
The differential fluxes of neutrinos and photons from the Galactic center 
can be easily calculated from the formula~\cite{bertone} for the case of 
Dirac DM:
\bea
 \frac{d F_i}{d E_i}(\psi,E)&=& \langle \sigma v \rangle {d N_i \over d E} {1 \over 16 \pi M_{\psi_D}^2}
 \int_{\text{line of sight}} d s \rho^2(r(s,\psi)),
\eea
where $i = \nu, \gamma$ and $s$ is the distance from the Earth in the angular direction $\psi$
from the line connecting the Earth and the Galactic center (GC).

The neutrinos from the GC can be detected at the superkamiokande (SK)
as the muon neutrios transform into the muons through 
the weak interactions in the rocks below the SK.
The neutrino-induced muon flux is written as
\bea
F_{\mu^+ \mu^-} = \int d E_{\nu_\mu} \frac{d F_{\nu_\mu}}{d E_{\nu_\mu}} f(E_{\nu_\mu}).
\eea
The function $f(E_{\nu_\mu})$ is the probability of a muon neutrino with 
energy $E_{\nu_\mu}$ transforming into muon with energy larger than $E_{\rm th}$,  and is given by~\cite{Hisano:2009}
\bea
 f(E_{\nu_\mu}) &=& \int_{E_{\rm th}}^{E_{\nu_\mu}} d E_\mu
  \left(
\frac{d \sigma_{\nu_\mu (\overline{\nu}_\mu)p \to \mu(\overline{\mu})X}}{d E_\mu} n_p^{(\rm rock)} 
+
\frac{d \sigma_{\nu_\mu (\overline{\nu}_\mu)n \to \mu(\overline{\mu})X}}{d E_\mu} n_n^{(\rm rock)} 
  \right)
R(E_\mu, E_{\rm th}),
\eea
where ${d \sigma_{\nu_\mu (\overline{\nu}_\mu)p(n) \to \mu(\overline{\mu})X}}/{d E_\mu}$ is
the scattering cross section of a neutrino with proton (neutron) to create
a muon with energy $E_\mu$~\cite{Barger:2007}:
\bea
\frac{d \sigma_{\nu_\mu (\overline{\nu}_\mu)p \to \mu(\overline{\mu})X}}{d E_\mu} 
&=& 
\frac{2 m_p G_F^2}{\pi} \left(0.21 + 0.29 \frac{E_\mu^2}{E_{\nu_\mu}^2}\right), \nl 
\frac{d \sigma_{\nu_\mu (\overline{\nu}_\mu)n \to \mu(\overline{\mu})X}}{d E_\mu} 
&=& 
\frac{2 m_p G_F^2}{\pi} \left(0.29 + 0.21 \frac{E_\mu^2}{E_{\nu_\mu}^2}\right).
\eea
For the number density of proton (neutron) in the rock, we use
$n_{p}^{(\rm rock)} = n_{n}^{(\rm rock)} = 2.65 N_A/2 \,{\rm cm}^{-3}$ 
($N_A = 6.022 \times 10^{23}$). 
$R(E_\mu, E_{\rm th})$ is the distance a muon with $E_\mu$ can travel inside the rock before losing energy below $E_{\rm th}$, and 
is fitted to be~\cite{Hisano:2009}
\bea
 R(E_\mu, E_{\rm th}=10 {\rm GeV}) = 10^{a + b y + c y^2} (\rm km),
\eea
where $y=\log_{10}(E_\mu/1 {\rm GeV}), a=-3.29186, b =1.52594$, and $c=-0.147224$.

\FIGURE{
\epsfig{file=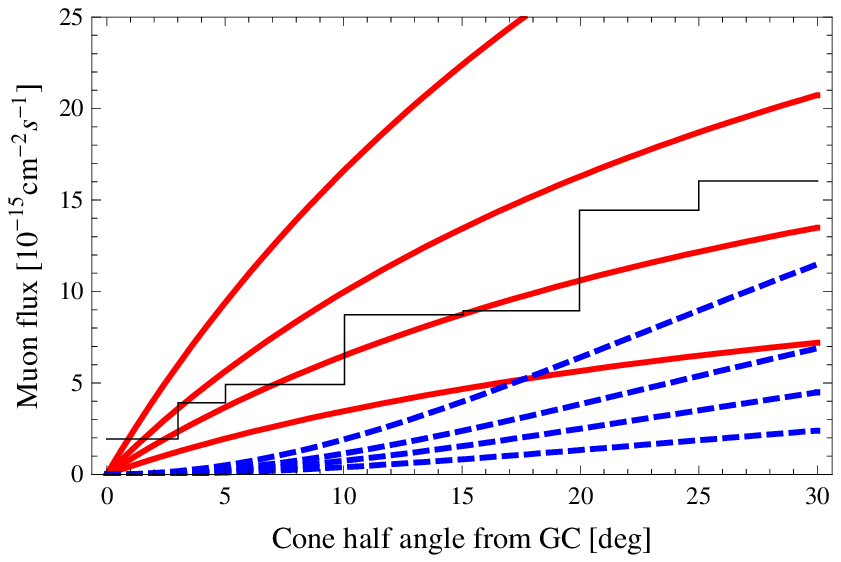,width=12cm,height=9cm}
\caption{
Thick solid red curves (thick dashed blue curves) are predictions of 
the neutrino-induced up-going muon flux from the annihilation of 
dark matter with masses 3, 2, 1.5, 1 TeV from above,
for the NFW (isothermal) dark matter profile.
The thin solid line is the superkamiokande bound.
}
\label{fig:muon_flux}
}

Fig.~\ref{fig:muon_flux} shows the predictions for the neutrino-induced 
muon flux for the DM masses $M_{\psi_D} = 3, 2, 1.5, 1$ TeV 
(from above). We obtained the annihilation cross section
in such a way that each DM mass fits the PAMELA, Fermi and HESS data 
as described above.
We used the NFW (solid red curves) and the isothermal (dashed blue curves) profiles for the plot. We can see that the 3 TeV DM is already ruled out by the SK bound because it needs too large BF. 
The 2 TeV DM which fits the CR data best is only marginally allowed. 
The lower DMs are allowed with the NFW profile.
However, if the isothermal profile is used, all the DM are allowed because this profile is flat near the Galactic center and the neutrinos are not much produced.

Fig.~\ref{fig:photon} shows the predictions for the gamma-ray flux from the Galactic 
center ($0.1^\circ$ region from the GC)~\cite{HESS_GC} and the Galactic Center 
ridge ($|b|<0.3^\circ, |l|<0.8^\circ$)~\cite{HESS_GR}.
We can see that the constraints on the DM annihilation for the NFW profile become more severe 
than in the neutrino case.
That is the NFW predicts too much gamma-ray, exceeding even the current data for the massive
DM. However, if more flat profile like the isothermal profile is used, the predictions are
below the current data.  

\FIGURE{
\epsfig{file=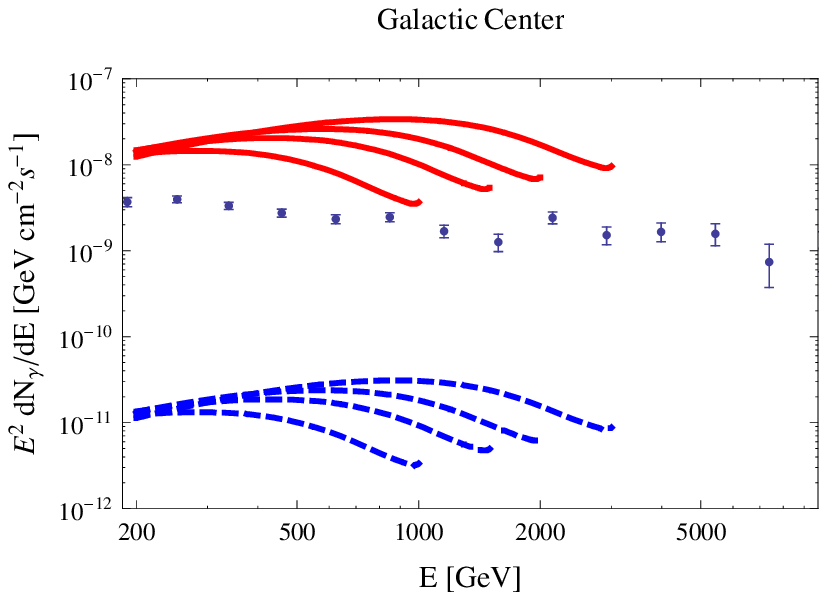,width=7cm,height=5cm}
\epsfig{file=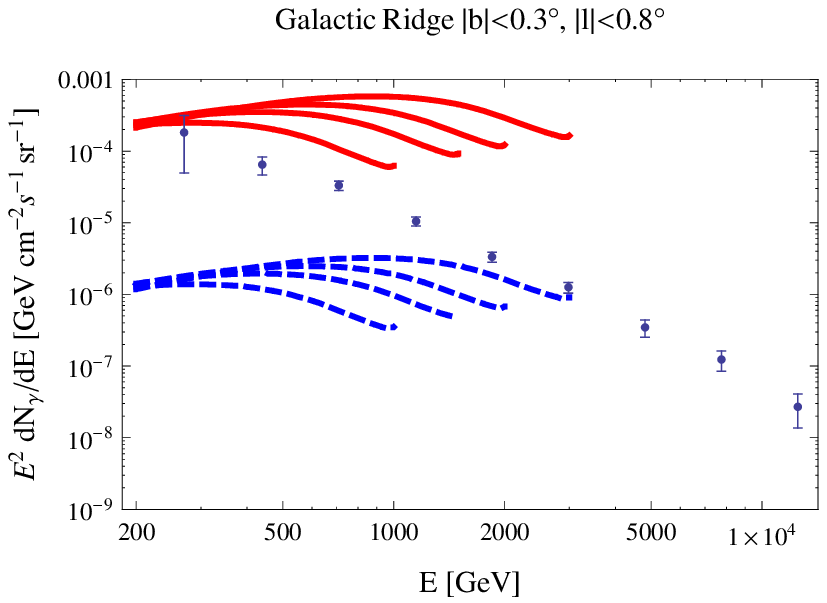,width=7cm,height=5cm}
\caption{
The gamma ray flux from the GC (left panel) and GC ridge (right panel).
Thick solid red curves (thick dashed blue curves) are predictions of the 
gamma ray flux from the annihilation of 
dark matter with masses 3, 2, 1.5, 1 TeV from above,
for the NFW (isothermal) dark matter profile.
}
\label{fig:photon}
}

\section{Collider Signatures}
\label{sec:collider}

New particles in this model are $Z^{'}$, $s$ (the modulus of $\phi$) 
and $\psi_D$. $Z^{'}$ couples only to muon, tau or their neutrinos, or 
the $U(1)_{L_\mu - L_\tau}$ charged dark matter. 
The new scalar $s$ can mix with the SM Higgs boson $h_{\rm SM}$, 
affecting the standard Higgs phenomenology.  

Let us discuss first the decay of $Z^{'}$ gauge boson and 
its productions at various colliders.
In the broken phase with $M_{Z^{'}} \neq 0$,
$Z^{'}$ can decay through the following channels:
\[
Z^{'}  \rightarrow  \mu^+ \mu^- , \tau^+ \tau^- ,
\nu_\alpha \bar{\nu}_\alpha \ ({\rm with} \ \alpha =
\mu \ {\rm or} \ \tau) , \
\psi_D \overline{\psi}_D \ ,
\]
if they are kinematically allowed. Since these decays occur through
$U(1)_{L_\mu - L_\tau}$ gauge interaction, the branching ratios are 
completely fixed once particle masses are specified. In particular,
\[
\Gamma ( Z^{'} \rightarrow \mu^+ \mu^- ) =
\Gamma ( Z^{'} \rightarrow \tau^+ \tau^- ) =
2 \Gamma ( Z^{'} \rightarrow \nu_\mu \bar{\nu}_\mu ) =
2 \Gamma ( Z^{'} \rightarrow \nu_\tau \bar{\nu}_\tau )
= \Gamma ( Z^{'} \rightarrow \psi_D \bar{\psi}_D )
\]
if $M_{Z^{'}} \gg m_\mu , m_\tau , M_{\rm DM}$.  The total decay
rate of $Z^{'}$ is approximately given by
\[
\Gamma_{\rm tot} (Z^{'}) = {\alpha^{'}\over 3}~ M_{Z^{'}} \times
4 (3)
\approx {4({\rm or} \ 3) \over 3}~{\rm GeV}~
\left( \alpha^{'} \over 10^{-2} \right)
~\left( {M_{Z^{'}} \over 100 {\rm GeV}} \right)
\]
if the channel $Z^{'} \rightarrow \psi_D \bar{\psi}_D$ is open (or closed).
Therefore $Z^{'}$ will decay immediately inside the detector for a
reasonable range of $\alpha^{'}$ and $M_{Z^{'}}$.

$Z'$ can be produced at a muon collider as resonances in the
$\mu \mu$ or $\tau\tau$ channel ~\cite{baek_he_ko} via
\[
\mu^+\mu^- \rightarrow Z^{'*} \rightarrow \mu^+\mu^- (\tau^+ \tau^-) .
\]
The LHC can also observe the $Z'$ which gives the right amount of
the relic density as can be seen in Fig.~1.
Its signal is the excess of multi-muon (tau) events without the excess of multi-$e$ events.

The dominant mechanisms of $Z^{'}$ productions at available 
colliders are 
\begin{eqnarray*}
q \bar{q} \ ({\rm or} \ e^+ e^-)
& \rightarrow & \gamma^* , Z^* \rightarrow \mu^+ \mu^- Z^{'} ,
\tau^+ \tau^- Z^{'} 
\\
& \rightarrow &  Z^* \rightarrow \nu_\mu \bar{\nu}_\mu Z^{'} ,
\nu_\tau \bar{\nu}_\tau Z^{'} 
\end{eqnarray*}
There are also vector boson fusion processes such as
\begin{eqnarray*}
W^+ W^- & \rightarrow & \nu_{\mu} \bar{\nu}_{\mu} Z^{'}  \ \
( \rm{or} \    \mu^+ \mu^-  Z^{'}) , \ \ \ {\rm etc.}
\\
Z^0 Z^0  & \rightarrow & \nu_{\mu} \bar{\nu}_{\mu} Z^{'}  \ \
( \rm{or} \    \mu^+ \mu^-  Z^{'}) , \ \ \ {\rm etc.}
\\
W^+ Z^0 & \rightarrow & \nu_{\mu} \bar{\mu} Z^{'}  \ \
( \rm{or} \    \mu^+ \mu^-  Z^{'}) , \ \ \ {\rm etc.}
\end{eqnarray*}
and the channels with $\mu \rightarrow \tau$. We will ignore the
vector boson fusion channels in this paper, since their contributions 
are expected to be subdominant to the $q\bar{q}$ or $e^+ e^-$ annihilations.

In Fig.~1, we present the $Z^{'}$ production cross sections at B factories, 
$Z^0$ pole, LEP(2), Tevatron and LHC. We find that the light $M_{Z^{'}}$ region  that can accommodate the muon $(g-2)_\mu$ is almost excluded 
by the current collider data.  The remaining region can be covered 
at the LHC with high integrated luminosity $\gtrsim$ 50 fb$^{-1}$.

The signatures of $Z^{'}$ will be an $s -$channel resonance in the dimuon 
invariant mass spectrum, or its deviation from the SM predictions as in 
Drell-Yan production of the muon pair.
Therefore one could expect that the number of multi-muon events at colliders
is enhanced compared with the SM predictions.
The $e^+ e^-$ channel will be diluted compared with the $\mu^+ \mu^-$ channel,
since the $e^+ e^-$ final state in the $Z^{'}$ decay can appear 
only through the $Z^{'}$decay into a tau pair and 
$\tau \rightarrow e \nu\bar{\nu}$ in this model.

Now let us discuss the Higgs phenomenology in our model. In general, 
there can be a mixing between the SM Higgs boson $h_{\rm SM}$ and 
a new scalar $s$ (the modulus of $\phi$) due to the $\lambda_{H\phi}$ 
coupling in (\ref{eq:L_new}). As a consequence, the Higgs searches 
at colliders can be quite different from those of the SM.
For example, one can imagine 
\[
g g \rightarrow h_{\rm SM}^* \rightarrow s^* \rightarrow
Z^{'} Z^{'} ,
\]
followed by $Z^{'} \rightarrow \mu^+ \mu^- , \tau^+ \tau^- ,
\nu_\mu \bar{\nu}_\mu , \nu_\tau \bar{\nu}_\tau , 
\psi_D \overline{\psi}_D$.
This makes an additional contribution to the $Z^{'}$ production 
at the LHC.  However, we did not include this $Z^{'}$ pair production 
through gluon fusion in Fig.~1 for simplicity, since it depends on the unknown free parameter $\lambda_{H\phi}$, and thus is more model dependent.
In any case, the generic collider signatures of the new $Z^{'}$ are
the excess of multi-muon or tau events, compared with the SM.
It is strongly desirable to search for $\mu \mu\mu\mu$,
$\tau \tau\tau\tau$, or $\mu\mu\tau\tau$ or large missing $E_T$ 
from $\mu\mu \nu\nu$ events at LEP, LEP2, Tevatron and at the LHC.

We can introduce an angle $\beta$ so that the ratio of two scalar VEV's 
is given by $\tan\beta = v_\phi/v_{h_{\rm SM}}$, and an angle 
$\alpha$ that parametrizes the mixing of $h_{\rm SM}$ and $s$: 
\begin{eqnarray}
h_{\rm SM} &=& H_1 \cos\alpha - H_2 \sin\alpha, \nonumber\\
s &=& H_1 \sin\alpha + H_2 \cos\alpha,
\end{eqnarray}
where $H_{1(2)}$ denotes the lighter (heavier) mass eigenstate of two scalars.

\begin{figure}
  \subfigure{\includegraphics[width=70mm]{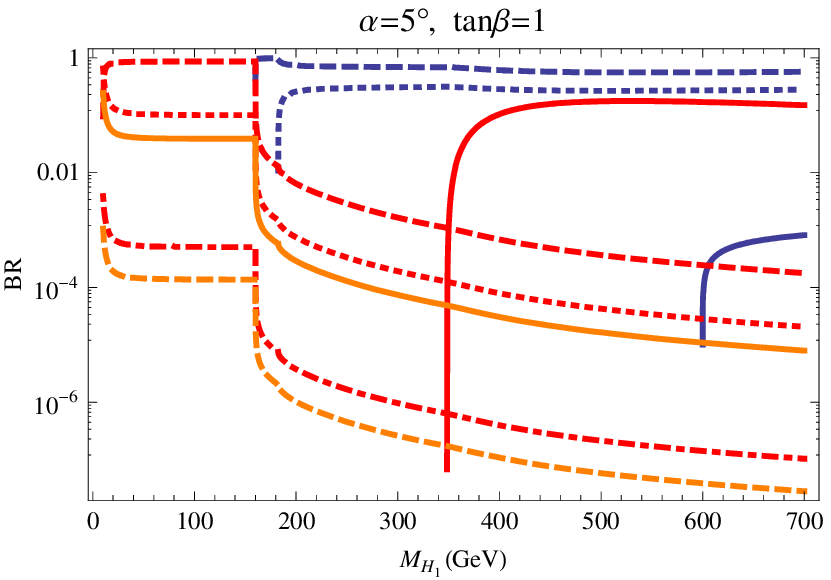}}
  \subfigure{\includegraphics[width=70mm]{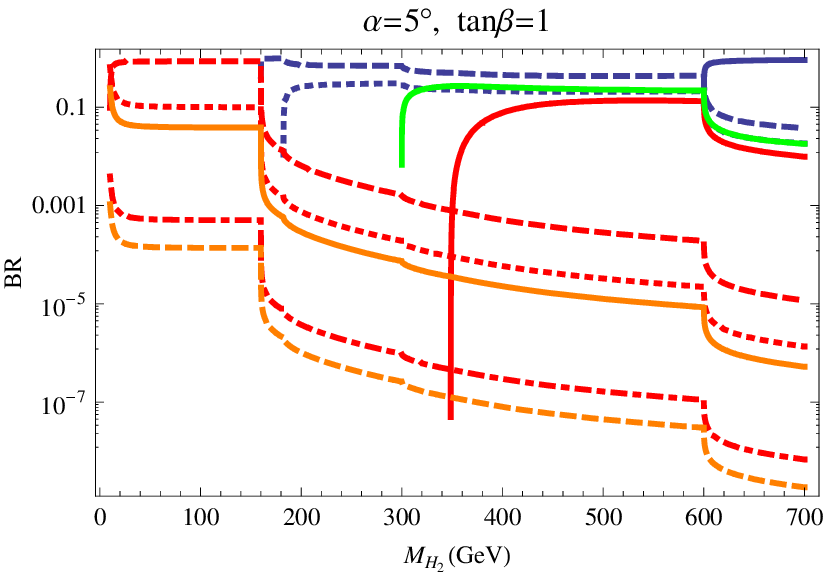}}
\\
  \subfigure{\includegraphics[width=70mm]{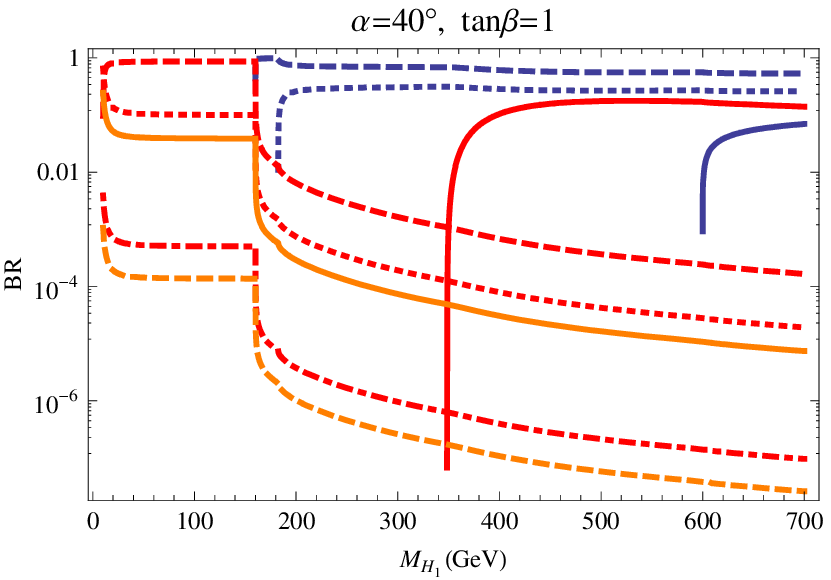}}
  \subfigure{\includegraphics[width=70mm]{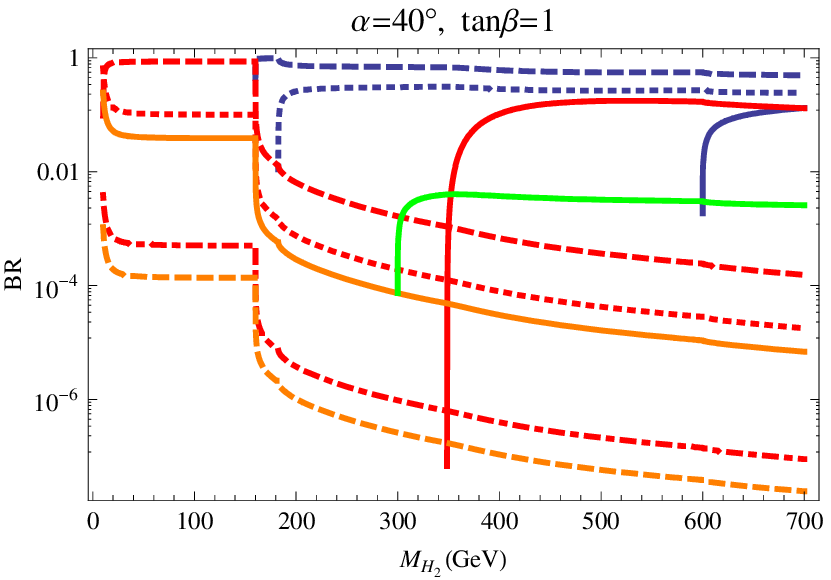}}
\\
  \subfigure{\includegraphics[width=70mm]{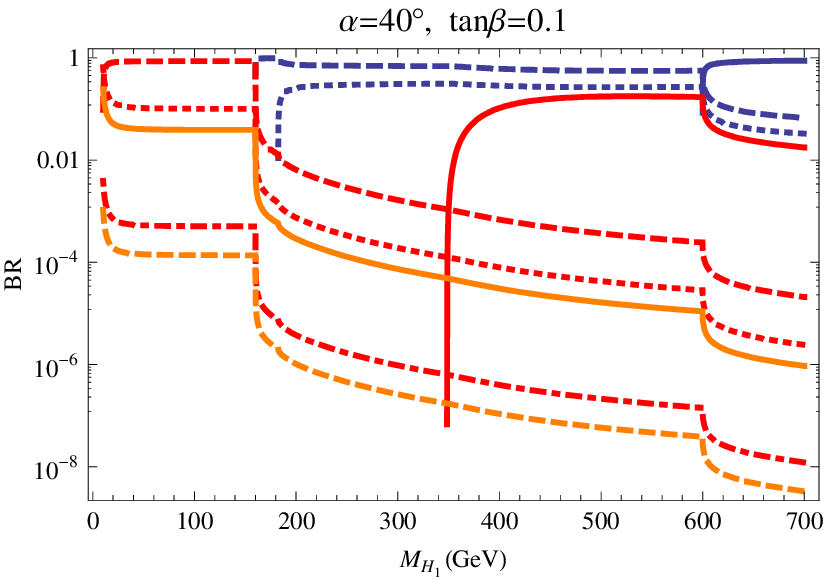}}
  \subfigure{\includegraphics[width=70mm]{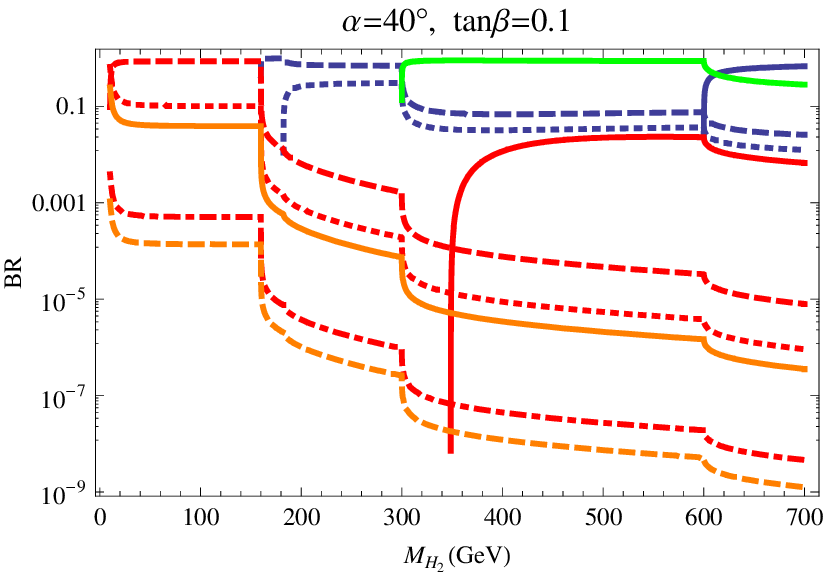}}
\caption{In the left (right) column are shown the branching ratios of the
lighter (heavier) Higgs $H_{1(2)}$ into two
particles in the final states:
$t\bar{t}$ (solid in red), $b\bar{b}$ (dashed red),
$c\bar{c}$ (dotted red), $s\bar{s}$ (dot-dashed red),
$\tau \bar{\tau}$ (solid orange), $\mu\bar{\mu}$ (dashed orange),
$WW$ (dashed blue), $ZZ$ (dotted blue) and $Z^{'} Z^{'}$ (solid blue)
for difference values of the mixing angle $\alpha$ and $\tan\beta$.
We fixed $M_{Z'} = 300$ GeV. We also fixed $M_{H_2}= 700$ GeV ($M_{H_1}=150$ GeV) for the plots
of the left (right) column.
}
  \label{fig:Hdecay}
\end{figure}

In Fig.~\ref{fig:Hdecay}, we show the branching ratios (BRs) of the 
two-body decay modes  of $H_{1,2}$ for $M_{Z'} = 300$ GeV.  
We have fixed $M_{H_2}= 700$ GeV ($M_{H_1}=150$ GeV) for the plots
of the $H_1$ ($H_2$) decay [ the left (right) column ].
Note that the modes $H_{1,2} \rightarrow Z' Z'$ (solid blue) and 
$H_2 \rightarrow H_1 H_1$ (solid green),  which are absent in the 
SM Higgs decay, can dominate for large $\alpha$ and small $\tan\beta$.
If $H_i$'s and $Z^{'}$ are heavy enough compared with the CDM 
in our model,  a decay $H_i \rightarrow Z^{'} Z^{'}$ followed by one or 
both of the $Z^{'}$ decaying into a pair of CDM or a pair of neutrinos 
could occur. Therefore the Higgs could have somewhat large invisible 
branching ratio, compared with the SM Higgs boson.  Therefore 
Higgs signatures at the Tevatron or the LHC could be quite exotic .

\section{Conclusions}
Recent possible anomalies in the cosmic ray data reported by 
PAMELA, Fermi-LAT and HESS Collaborations may be due to 
astrophysical origins such as pulsars.
However, it is also tantalizing to consider them as the first hint 
for the existence of weakly interacting cold dark matter.
In this paper, we considered a leptophilic CDM model with extra 
$U (1)_{L_{\mu} - L_{\tau}}$ gauge symmetry which is one of the anomaly
free global symmetry in the SM. We have introduced a new complex
scalar $\phi$ and Dirac fermion $\psi_D$ which are charged under 
the new $U(1)_{L_{\mu} - L_{\tau}}$.
The $U(1)_{L_{\mu} - L_{\tau}}$ charged Dirac fermion $\psi_D$ 
can be a good CDM that might explain the positron excess reported 
by HEAT, PAMELA and FERMI,  without producing excess in antiproton 
flux as observed by PAMELA.
This model is constrained by the muon $(g-2)_\mu$ and collider searches 
for a vector boson decaying into $\mu^+ \mu^-$ at the Tevatron, LEP(2) and $B$ factories. 
The collider constraints favors $\psi_{\rm DM}$ heavier than $\sim$ 100 GeV. We calculated the relic density of the CDM with these 
constraints, and still find that the thermal relic density could be easily
within the WMAP range. We also considered the production cross section
of the new gauge boson $Z^{'}$ at the LHC, which could be 1 fb --1000 fb.
The new gauge boson $Z^{'}$ will decay into $\mu\bar{\mu}$, 
$\tau\bar{\tau}$, their neutrino partners or even to a pair of CDM's.
Therefore the final states will be rich in muons or taus, or missing $E_T$. 
Most parameter space of this model is within the discovery range at 
the LHC with enough integrated luminosity $\gtrsim 50 \ {\rm fb}^{-1}$. 
It is remained to be seen whether there are excess in the multimuon or 
multitau events at the Tevatron or at the LHC.

\section{Acknowledgements}

We are grateful to Seong Chan Park and Stefano Scopel for
useful discussions. 
The work was supported in part by the Korea Research Foundation Grant
funded by the Korean Government (MOEHRD) No. KRF-2007-359-C00009 (SB).

\end{document}